\documentclass[journal,10pt,letterpaper]{IEEEtran}
\pdfoutput=1
\IEEEoverridecommandlockouts
\usepackage{amsmath,amsfonts}
\usepackage{xcolor}
\usepackage{algorithmic}
\usepackage{algorithm}
\usepackage{array}
\usepackage{textcomp}
\usepackage{stfloats}
\usepackage{url}
\usepackage{verbatim}
\usepackage{graphicx}
\usepackage{cite}
\usepackage{subcaption}
\usepackage{booktabs}
\usepackage{setspace}
\usepackage{subfig}
\usepackage{afterpage}
\usepackage[letterpaper, left=1.01in, right=1.02in, top=1in, bottom=1.05in]{geometry}
\usepackage{placeins}

\newcommand{\fref}[1]{Fig.~\ref{#1}}
\newcommand{\sref}[1]{Section~\ref{#1}}
\newcommand{\aref}[1]{Algorithm~\ref{#1}}
\newcommand{\tref}[1]{Table~\ref{#1}}

\begin{document}

\title{Multi-Modal Fusion-Based Multi-Task Semantic Communication System}

\author{Zengle Zhu,
        Rongqing Zhang,~\IEEEmembership{Member,~IEEE}
        Xiang Cheng,~\IEEEmembership{Fellow,~IEEE}
        Liuqing Yang,~\IEEEmembership{Fellow,~IEEE}

\thanks{
Zengle Zhu is with the Intelligent Transportation Thrust, Hong Kong University of Science and Technology (Guangzhou), Guangzhou 510000, China (e-mail: zzhu622@connect.hkust-gz.edu.cn).  
          
Rongqing Zhang is with the School of Software Engineering, Tongji University, Shanghai 200092, China (e-mail: rongqingz@tongji.edu.cn).
          
Xiang Cheng is with the School of Electronics, Peking University, Beijing 100871, China (e-mail: xiangcheng@pku.edu.cn).
          
Liuqing Yang is with the Internet of Things Thrust \& Intelligent Transportation Thrust, Hong Kong University of Science and Technology (Guangzhou), Guangzhou 510000, China, and also with the Department of Electronic and Computer Engineering, Hong Kong University of Science and Technology, Hong Kong SAR, China (e-mail: lqyang@ust.hk).
}}


\thispagestyle{empty}
\maketitle
\thispagestyle{empty}
\begin{abstract}
In recent years, there has been significant progress in semantic communication systems empowered by deep learning techniques. It has greatly improved the efficiency of information transmission. Nevertheless, traditional semantic communication models still face challenges, particularly due to their single-task and single-modal orientation. Many of these models are designed for specific tasks, which may result in limitations when applied to multi-task communication systems. Moreover, these models often overlook the correlations among different modal data in multi-modal tasks. It leads to an incomplete understanding of complex information, causing increased communication overhead and diminished performance. To address these problems, we propose a multi-modal fusion-based multi-task semantic communication (MFMSC) framework. In contrast to traditional semantic communication approaches, MFMSC can effectively handle various tasks across multiple modalities. Furthermore, we design a fusion module based on Bidirectional Encoder Representations from Transformers (BERT) for multi-modal semantic information fusion. By leveraging the powerful semantic understanding capabilities and self-attention mechanism of BERT, we achieve effective fusion of semantic information from different modalities. We compare our model with multiple benchmarks. Simulation results show that MFMSC outperforms these models in terms of both performance and communication overhead.
\end{abstract}


\section{Introduction} 
\IEEEPARstart{W}{ith} the rapid development of communication technology, data rates in wireless communications are rapidly approaching Shannon's capacity limit\cite{shannon1948mathematical}. The continuously increasing demand for communication causes the explosion of wireless data traffic, placing a heavy burden on the current infrastructure of communication systems\cite{xie2021deep}. Researchers have explored innovative approaches to optimize the efficiency of communication systems\cite{wang2015multi, han2018propagation, xie2021deep}. Among them, semantic communications\cite{xie2021deep} have emerged as a promising solution. Different from traditional communication systems that primarily emphasize the accurate transmission of bit streams, semantic communications only transmit task-related semantic information from the source data. By focusing on the semantic features, semantic communication systems can achieve superior performance while transmitting less data. Many works\cite{weng2021semantic, dong2022semantic, xie2021deep, wang2022wireless} have shown its outstanding capabilities, especially in adverse channel conditions, making it an important technology for the future of communications. 

At present, most existing works on semantic communications focus on a specific task\cite{weng2021semantic, dong2022semantic, xie2021deep}. However, in practical scenarios, we usually need to handle multiple tasks. This poses a challenge for single-task semantic communication systems. The feasible approaches are to continuously update the model or store multiple task-specific models. Continuous model updating requires a significant amount of computational resources, whereas storing multiple task-specific models increases the complexity of systems and storage resource consumption\cite{zhang2022unified}. Another noteworthy limitation of existing models is that they overlook the significance of correlations among multiple modalities. In the context of multi-modal tasks, models such as those proposed by Zhang \emph{et al}. \cite{zhang2022unified}, Wang \emph{et al}. \cite{wang2023cooperative}, and Xie \emph{et al}. \cite{xie2021task} fail to capture the correlations among various modal data. This means that there is a lack of clear and accurate knowledge of the target task. In multi-modal tasks, different types of data (such as text, image, speech, and video) often contain complementary information. Understanding the correlations among these modalities is critical to perform tasks accurately. Moreover, multi-modal data will increase communication overhead. These two reasons hinder the development of semantic communication technologies. Therefore, we need to develop a multi-modal fusion-based multi-task semantic communication framework which can adapt to multi-task scenarios and effectively utilize the complementary information provided by different modalities. At the same time, this is also expected to enhance performance and wider usability in real-world scenarios. However, the application of this framework has the following challenges:

\subsubsection{Communication latency and bandwidth overhead} 
The transmission of information across various modal data often requires greater bandwidth. It leads to higher communication latency, significantly impacting the real-time responsiveness.

\subsubsection{Semantic fusion among heterogeneous data} 
Different data modalities encompass a wide array of data types. It is important to mine the complex semantic correlations among heterogeneous data. How to effectively fuse multi-modal data is a complex challenge.

\subsubsection{Complexity in multi-task and multi-modal framework} 
The addition of tasks and modalities escalates the complexity of communication systems, which may influence performance on tasks. As the complexity increases, it is crucial to ensure that task performance is not affected.

Inspired by prior research, we propose a multi-modal fusion-based multi-task semantic communication (MFMSC) framework. Unlike traditional single-task models or those ignore multi-modal semantic complementarity, our model can not only handle different tasks, but also exploit the semantic relationships among different modal data. The effectiveness of the proposed framework is verified by extensive experiments. Simulation results show that our framework outperforms multiple benchmarks in terms of communication overhead and task performance. The main contributions are summarized as follows.

\begin{itemize}
  \item We construct an innovative semantic communication architecture to mitigate the performance impact of growing tasks and data modalities. Specifically, we design dedicated semantic encoders for each data modality, enabling more accurate capture of modality-specific features. Moreover, we introduce task embeddings for each task, which can effectively prevent mutual interference among different tasks. This design ensures that the designed model can still maintain high performance along with increasing tasks and data modalities.
  \item In order to solve the inherent challenges in multi-modal task scenarios, we further design a novel and efficient fusion module based on Bidirectional Encoder Representations from Transformers (BERT). It effectively fuses multi-modal semantic information extracted by the semantic encoder corresponding to each modality, resulting in a significant improvement in task performance. Moreover, the fusion module can greatly reduce communication overhead by eliminating the redundancy. It effectively solves the delay problem and improves the efficiency of our framework
  \item In this work, we select 8 datasets and conduct extensive experiments. We compare various baselines and existing methods. Experimental results show that the MFMSC framework has superior performance, consistently achieving or approaching the state-of-the-art (SOTA) level across various tasks. Especially in multi-modal tasks, our framework improves performance by about 10\% compared to those do not consider multi-modal fusion, and greatly reduces the amount of data transmitted. These results demonstrate the application prospects of our framework.
\end{itemize}

The rest of this paper is organized as follows: \sref{work} introduces the related works and explores previous reasearch in related fields. MFMSC system is presented and a corresponding problem is formulated in \sref{problem}. \sref{framework} details the architecture of MFMSC framework. \sref{results} shows the performance of MFMSC by simulation results. \sref{conclusion} concludes this paper.

\section{Related Works}
\label{work}
\subsection{Semantic Communications}
According to Weaver\cite{weaver1953recent}, communications can be categorized into three levels: the technical level, which ensures the accuracy of bit sequence transmission; the semantic level, which ensures transmitted symbols accurately convey the desired meaning; and the effectiveness level, which ensures the received meaning affect conduct in the desired way. In the contemporary field of wireless communications, data rates are approaching Shannon's capacity limit. Traditional data transmission methods established at the technical level often require high communication overhead, especially when dealing with large datasets or multi-modal data\cite{yang2022semantic}. Therefore, we need a more efficient communication method. Building upon Weaver's seminal categorization\cite{weaver1953recent}, many studies \cite{weng2021semantic, xie2021deep, dong2022semantic, wang2022wireless} are dedicated to studying a new paradigm at the semantic level, namely semantic communications.

Semantic communication systems show great promise. They interpret information at the semantic level rather than mere bit sequences\cite{xie2021deep}, which aims to achieve more intelligent, efficient, and reliable data transmission. The benefits of semantic communications mainly lie in reducing communication overhead and improving downstream task performance. They eliminate redundant content and focus on the meaningful information for downstream tasks, thus preventing the impact of irrelevant information on task performance. On the other hand, by transmitting task-related information, the bandwidth and communication latency are reduced. 

The opening work on semantic communications is DeepSC\cite{xie2021deep}, which is developed for text reconstruction. This research applies Transformer\cite{vaswani2017attention} to communications. Due to the self-attention mechanism, the performance of DeepSC far exceed traditional communication methods. As more and more semantic communication models are proposed\cite{weng2021semantic, dong2022semantic, wang2022wireless}, semantic communications have expanded beyond text transmission and are now active in the transmission of other modal data such as speech, image, and video.

However, existing works mainly concentrate on single-task or single-modal scenarios. Due to the growing diversity of data, information is no longer limited to a single format or target. Image, text, speech, and video data often coexist, and we need to use them together to perform various tasks. Moreover, current research often overlooks the correlations among multi-modal data. This reduces the ability of models to understand tasks. By solving the challenges posed by multiple modalities and multiple tasks, communication systems can represent semantic information more deeply and accurately. In this paper, we delve into these shortcomings, aiming to bridge the gap between existing semantic communication research and the demands of real-world applications, thereby creating a more comprehensive and effective multi-modal and multi-task semantic communication framework.

\subsection{Deep Learning for Multiple Tasks and Multiple Modalities}
Recent research has made significant strides in multi-task models\cite{liu2019end, xu2022mtformer}. Various techniques like soft sharing\cite{misra2016cross} and hard sharing\cite{liu2019multi} have been explored to enhance the performance of individual tasks through cross-task knowledge transfer. Hard sharing is the most widely used sharing mechanism. It embeds the data representations of multiple tasks into the same semantic space, and then uses a task-specific layer to extract task-specific representations. Hard sharing is suitable for processing tasks with strong correlations, but it often performs poorly when encountering tasks that are less related\cite{sun2020learning}. Soft sharing builds a unique neural network for each task. It is suitable for situations where the correlations among tasks are not strong\cite{sun2020learning} but increases storage overhead. In addition to the inherent shortcomings, these approaches often face challenges when applied to tasks involving multi-modal data.

In multi-modal tasks, the focus is on leveraging complementary information from different data modalities. Fusion techniques such as early fusion, intermediate fusion, and late fusion have shown promise\cite{atrey2010multimodal}. Many researchers use architectures such as Multilayer Perceptron (MLP) and Convolutional Neural Network (CNN)\cite{atrey2010multimodal} in multi-modal fusion, which has indeed made some progress. Nevertheless, these methods are difficult to adapt to the changing relationships among different data\cite{xue2023dynamic}. With the deepening of research, the attention mechanism\cite{vaswani2017attention} has gradually attracted researchers.

In various fields, the attention mechanism has shown excellent results. The Transformer\cite{vaswani2017attention} model improves the ability of text understanding, and models such as Vision Transformer\cite{dosovitskiy2020image}, Speech Transformer\cite{dong2018speech}, and Video Vision Transformer\cite{arnab2021vivit} introduce self-attention mechanism into image, speech and video processing, significantly improving the performance of corresponding tasks. These studies show that the attention mechanism exhibits outstanding performance and application potential, whether in different tasks or in different modal data.

Considering that tasks with the same modal data are usually more closely related, while tasks with different data modalities are relatively weakly related. Related tasks often have inter-dependence and perform better when solved in a joint framework\cite{akhtar2019multi}. Therefore, we propose to use a combination of hard sharing and soft sharing. We design independent semantic encoders for different modalities while utilizing shared channel encoder and channel decoder to achieve unified compression and recovery. In order to prevent negative transfer among unrelated tasks, we add additional task embeddings\cite{achille2019task2vec} for each task to enhance the distinction. To fully exploit the complementarity of multi-modal data and reduce the cost of transmission, we propose an innovative fusion module based on BERT\cite{devlin2018bert} to fuse multi-modal semantic information. In multi-modal tasks, the semantic features of each modality are extracted by corresponding semantic encoders. These features are concatenated as input sequences which are then semantically fused by the fusion module. This exploration improves the task performance and provides a new paradigm for multi-modal tasks in semantic communications.
\begin{figure*}[htbp]
  \centering
  \includegraphics[width=0.9\textwidth]{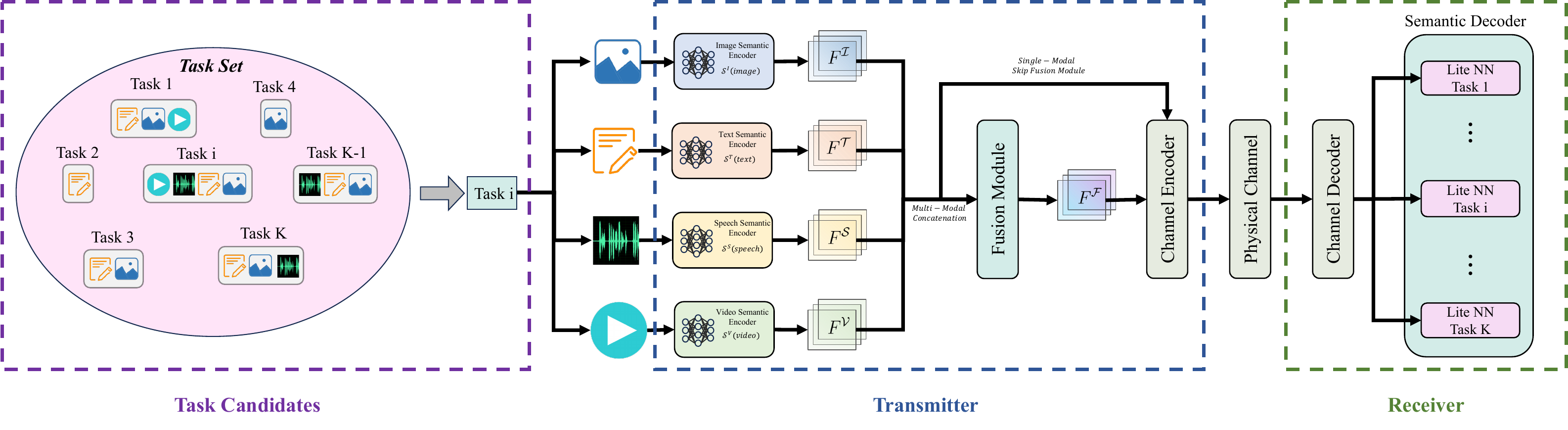}
  \vspace{1ex}
  \caption{The structure of our multi-modal fusion-based multi-task semantic communication system.}
  \label{fig:model}
\end{figure*}

\section{System Model and Problem Formulation}
\label{problem}
\subsection{Multi-Modal Fusion-Based Multi-Task Semantic Communication System}
Unlike traditional approaches that usually deal with single-modal data and single-task data, we consider multi-modal and multi-task scenarios that require processing different data modalities and performing multiple tasks simultaneously. Our multi-modal fusion-based multi-task semantic communication system is shown in \fref{fig:model}. The system encompasses a task set comprising numerous tasks, each contains various modal data. It is designed to support four modalities: image, text, speech, and video. In the system, three components play essential roles: transmitter, physical channel, and receiver. 

The transmitter consists of semantic encoder and channel encoder. In the semantic encoder, we build four distinct encoders, a text encoder, an image encoder, a speech encoder, and a video encoder. Each encoder is tailored to one modality. This architecture can better extract semantic information from different modal data. In addition, we design a BERT-based fusion module to fuse multi-modal semantic information. It takes the concatenation of different modal semantic information as input sequences, and further uses the self-attention mechanism to fuse them. This module enhances the processing capabilities for multi-modal tasks and significantly reduces the cost of data transmission. 

The primary function of the channel encoder is to compress the high-dimensional semantic information obtained from the semantic encoder. Utilizing neural network techniques, the channel encoder can effectively reduce the dimension of the data. This reduction eases bandwidth demands while maintaining the semantic integrity of the original semantic information.

The compressed semantic information will be transmitted to the physical channel. We mainly consider two channels, the Additive White Gaussian Noise (AWGN) channel and the Rayleigh fading channel. Both of them represent typical wireless channel environments. The AWGN channel is widely used to simulate an ideal wireless environment, and the Rayleigh fading channel is more suitable for modeling complex wireless environments, such as heavily built-up urban environments. By studying both wireless transmission environments, we can fully evaluate the performance of our system.

The semantic information reaches the receiver after passing through the physical channel. The receiver incorporates channel decoder and semantic decoder. The channel decoder based on deep learning is a key component of the semantic recovery. Its main responsibility is to reconstruct undistorted high-dimensional semantic information. The recovered semantic information is fed into the semantic decoder, which consists of a series of lite neural network task heads. Each task head is optimized for a specific task. The use of lite task heads can ensure fast response of the entire system.

\subsection{Problem Formulation}
In this work, we consider $K$ tasks, denoted by a set $\mathcal{K}=\{1,2, \cdots, K\}$. In the task set, there are multiple tasks $T = \{T_1, T_2, ..., T_K\}$, where $T_i$ represents the $i$-th task and $i  \in \mathcal{K}$. The modality of source data $D_i$ in $T_i$ is $M_{i} = \{M_1, M_2, ..., M_m\}$. We consider four modalities, namely $M_i \subseteq \{\mathcal{I}: image, \mathcal{T}: text, \mathcal{V}: video, \mathcal{S}: speech\}$.

We employ Deep Neural Network (DNN) to design both the transmitter and receiver. The transmitter consists of two parts, the semantic encoder and the channel encoder. They are designed to extract semantic information from $D_i$ and ensure the successful transmission over the physical channel. For task $T_i$, the encoded symbol stream $X_i$ can be represented as
\begin{equation} 
  \label{transmitter}
  X_i = C(S(D_i; \alpha_T); \beta_T)
\end{equation}
where $S(\cdot ; \alpha_T)$ is the semantic encoder network with the parameter set $\alpha_T$ and $C(\cdot; \beta_T)$ is the channel encoder network with the parameter set $\beta_T$. Then $X_i$ is transmitted over the physical channel. This process can be modeled as
\begin{equation}
  \label{channel}
  \hat{X_i} = HX_i + N
\end{equation}
where $H$ represents the channel matrix and $N \sim \mathcal{CN}(0, \sigma^2 I)$ is the Gaussian noise. 

The semantic information is sent to the receiver after passing through the physical channel. The channel decoder first restore semantic information and the semantic decoder processes the recovered semantic information to perform intelligent tasks. This process can be denoted as
\begin{equation}
  \label{receiver}
  \hat{Y_i} = S^{-1}(C^{-1}(\hat{X_i}; \beta_R); \alpha_R)
\end{equation}
where $S^{-1}(\cdot ; \alpha_R)$ is the semantic decoder network with the parameter set $\alpha_R$ and $C^{-1}(\cdot; \beta_R)$ is the channel decoder network with the parameter set $\beta_R$. $\hat{Y_i}$ is the predicted result of task $T_i$.

For task $T_i$, its performance metric function is given as $P_i(\hat{Y_i}, Y_i)$, where $Y_i$ represents the ground truth. Our goal is to maximize the metric function for every task
\begin{equation}
  \label{goal}
  \max_ {\substack{\{\alpha_T, \beta_T, \alpha_R, \beta_R\}}} P_i(\hat{Y_i}, Y_i), \quad \text{where} \quad i \in \mathcal{K}.
\end{equation}
In order to maximize the metric function of each task, it actually means that we need to minimize the loss function of them, which is given by
\begin{equation}
  \min_{\substack{\{\alpha_T, \beta_T, \alpha_R, \beta_R\}}} L_i(\hat{Y_i}, Y_i), \quad \text{where} \quad i \in \mathcal{K}.
\end{equation}
$L_i$ represents the loss function of $T_i$.

We strive to design a comprehensive semantic communication system to meet this objective. The system can adapt to the needs of different tasks and take advantage of the complementarity among various data modalities.

\subsection{Task Description}
To comprehensively analyze the system, we consider single-modal tasks of image, text, speech, and video as well as some multi-modal tasks.

\subsubsection{Text Tasks}
We consider sentiment analysis task and text reconstruction task. We utilize the GLUE-SST2 dataset for binary sentiment classification and the Europarl dataset for text reconstruction.
 
\subsubsection{Image Tasks}
For image modal, we focus on classification and reconstruction tasks, employing the CIFAR-10 dataset for both. The classification task is to identify the category, while the reconstruction task is to restore transmitted images.

\subsubsection{Speech Task}
The speech modal is centered on speech recognition using the LibriSpeech dataset, which provides rich spoken English recordings.

\subsubsection{Video Task}
We use the widely used HMDB51 dataset for our analysis. It serves to identify and classify human activities.

\subsubsection{Multi-modal Tasks}
To evaluate the effectiveness of the system in multi-modal tasks, we choose two multi-modal datasets: Visual Question Answering v2 (VQA v2) and Multimodal IMDb (MM-IMDb)\cite{arevalo2017gated}. The VQA v2 dataset contains images, question text, and corresponding answer text. Its task is to answer correctly based on the given image and question. MM-IMDb is a classification dataset of film and television short dramas. Its task is to perform genre classification based on the poster image and plot text.

We use distinct metric functions and loss functions for different task types. For classification tasks like text sentiment analysis, image classification, and video classification, we use accuracy to measure performance and cross entropy as loss function. We transform the VQA task into a classification task by treating each answer as a label, and we use the same loss and metric function. Since MM-IMDb is a multi-label dataset, we use cross entropy as loss function while using F1-score as the evaluation metric to balance precision and recall.

We also approach text reconstruction as a classification task and use cross entropy loss. We construct a vocabulary where words are assigned numerical values that function as class labels. In order to better evaluate the quality of reconstructed text, we use BLEU score as the metric function. It ranges from 0 to 1 and is used to assess the similarity  between source and reconstructed text. Higher scores denote greater similarity.

Similarly, in speech recognition, we build a vocabulary and convert this task into a classification task. We adopt connectionist temporal classification (CTC) loss specifically designed for speech recognition. And we utilize the word accuracy as the evaluation metric. Word accuracy measures the difference between the predicted transcriptions and the ground truth.

Lastly, image reconstruction is assessed using PSNR (Peak Signal-to-Noise Ratio), calculated as
\begin{equation}
  \mbox{PSNR} = 20 * log_{10}(\mbox{MAX}) - 10 * log_{10}(\mbox{MSE})
\end{equation} 
 where \mbox{MAX} represents the maximum pixel value of the image, and \mbox{MSE} denotes the mean squared error between the original and reconstructed images.

\section{Semantic Communication Transceiver Design}
\label{framework}

To address the issues of high communication latency, increased bandwidth demands, and the complexity of semantic fusion across different modal data, we propose a novel multi-modal fusion-based multi-task semantic communication framework. In this section, we elaborate the detailed design of the semantic transceiver.

\subsection{Semantic Encoder Design}
In our multi-modal fusion-based multi-task semantic communication framework, the semantic encoder plays a crucial role. In the design of semantic encoder, our innovation is mainly reflected in two places. First, we design specialized semantic encoders for different modalities, and each encoder aims to effectively extract corresponding semantic features. This is because we need to take into account the heterogeneity of data from different modalities. Using a single encoder for all modalities will ignore specific information. By designing dedicated encoders for each modal data, the framework can adapt to the complexity of different modalities. Tasks with the same modal data can also enhance knowledge sharing by using the same encoders. Second, considering the complementarity and redundancy of different modal semantic information in multi-modal tasks, we design a fusion module. Information across various modalities is typically complementary, with each providing distinct details. Through the fusion module, we can effectively integrate the semantic information of these different modalities to generate a more comprehensive feature representation. In addition, through multi-modal fusion, the amount of data in the communication can be significantly reduced. In the following, we detail the semantic encoder corresponding to each modality. 

\subsubsection{Image Semantic Encoder}
\begin{figure}[htbp]  
  \centering
  \includegraphics[scale=0.23]{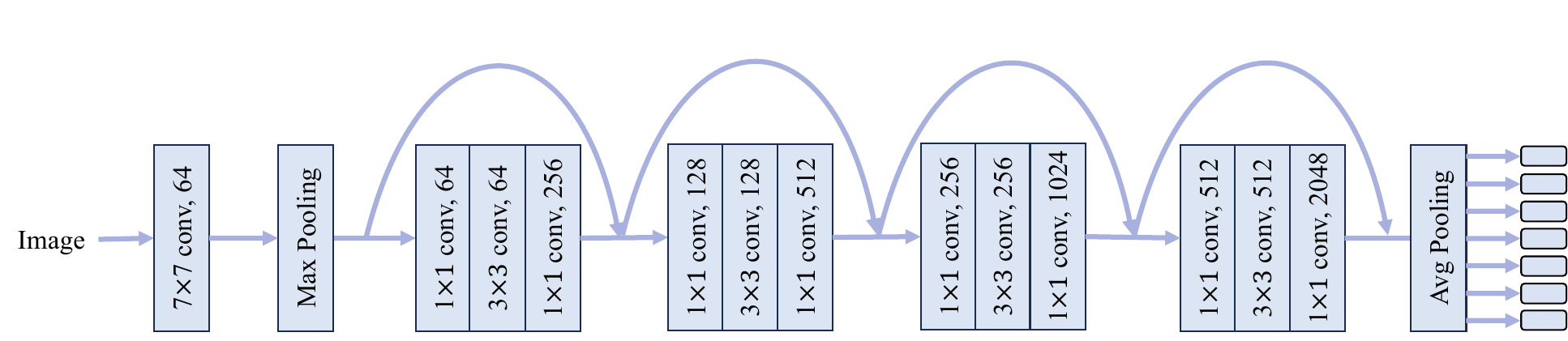}
  \caption{The structure of the image semantic encoder.}
  \label{fig:image_encoder}
\end{figure}

Images have complex features, including color, texture, and high-level semantic content. Due to this complexity, an effective image semantic encoder must be capable of capturing both low-level details and high-level information. Deepening the network enhances image feature extraction but also introduces the challenge of gradient vanishing. ResNet\cite{he2016deep} solves this problem through residual connections. Therefore, in terms of image data processing, we design our image semantic encoder based on the ResNet, which is shown in \fref{fig:image_encoder}. 

During preprocessing, we use data augmentation techniques to further improve the robustness of the image semantic encoder. Assuming that the input image data is $D^\mathcal{I} \in \mathbb{R}^{C \times H \times W}$, where $C$ represents the number of channels, $H$ is the height, and $W$ is the width of the image. The data passes through multiple convolutional layers and residual layers. The image semantic encoder has excellent semantic feature extraction capabilities by using residual layers. Consequently, we can obtain the semantic feature map $F^{\mathcal{I}} \in \mathbb{R}^{L_{\mathcal{I}} \times P}$ of the image data, where $L_\mathcal{I}$, $P$ represents the height and width of the feature map, respectively.

\subsubsection{Text Semantic Encoder}
\begin{figure}[htbp]  
  \centering
  \includegraphics[width=\linewidth,scale=1.]{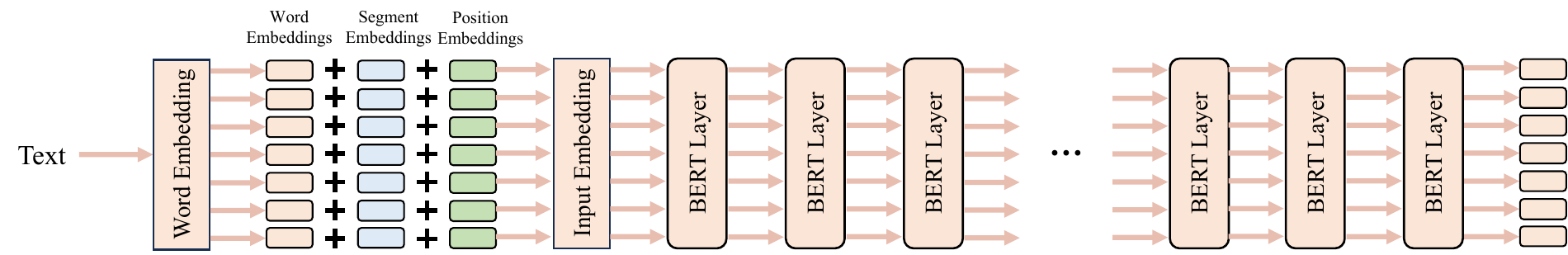}
  \caption{The structure of the text semantic encoder.}
  \label{fig:text_encoder}
\end{figure}

Text data has complex structures. In text, we need to consider grammar, polysemy, and contextual information. Complex text data requires a powerful model that can understand the meaning and relationships of words in different contexts. In order to better extract semantic information from text data, we build the text semantic encoder based on BERT\cite{devlin2018bert}, which is demonstrated in \fref{fig:text_encoder}. Unlike previous language models, BERT takes into account both the context of the text and the semantic information of each word, enabling a comprehensive grasp of sentence structure and meaning to extract precise semantic features.

To process the input text data $D^\mathcal{T}$, which is composed of sequences of words, our text semantic encoder employs the following steps. The first is tokenization. A tokenizer is used to decompose text into tokens. Assuming that $D^\mathcal{T}$ contains $N_\mathcal{T}$ words, which is denoted as $D^\mathcal{T} = [w_1, w_2, w_3, ..., w_{N_T}]$. The tokenizer maps words to their corresponding tokens. These tokens are then transformed into word embeddings, which serve as vector representations that capture the meaning and context of each token. The embedding vectors can be represented as $E^\mathcal{T}_w = [e_1, e_2, e_3, ..., e_{N_T}] \in \mathbb{R}^{N_T \times P}$. $P$ denotes the dimension of embedding vectors. In actual training  and testing process, the number of words in each batch is different. To align text data, vectors are truncated or zero-padded as necessary to ensure uniform dimensionality of $\mathbb{R}^{L_{\mathcal{T}} \times P}$ for each data instance.

Subsequently, we enhances the representation of each token by adding segment embeddings $E^\mathcal{T}_s \in \mathbb{R}^{L_{\mathcal{T}} \times P}$ and position embeddings $E^\mathcal{T}_p \in \mathbb{R}^{L_{\mathcal{T}} \times P}$, which is given by
\begin{equation}
  E^\mathcal{T}_{in} = E^\mathcal{T}_w + E^\mathcal{T}_p + E^\mathcal{T}_s
\end{equation}
where $E^\mathcal{T}_{in} \in \mathbb{R}^{L_{\mathcal{T}} \times P}$ is the input embeddings. Segment embeddings help our encoder distinguish different sentences in the same input, while position embeddings provide the information about token order. These embeddings are all learnable parameters.

Following this, the text semantic encoder processes the $E^\mathcal{T}_{in}$ through multiple layers, each comprising multi-head self-attention mechanism and feedforward neural network. The result is an extraction of semantic features $F^{\mathcal{T}} \in \mathbb{R}^{L_{\mathcal{T}} \times P}$, which contain rich contextual information.

\subsubsection{Speech Semantic Encoder}
\begin{figure}[htbp]  
  \centering
  \includegraphics[scale=0.25]{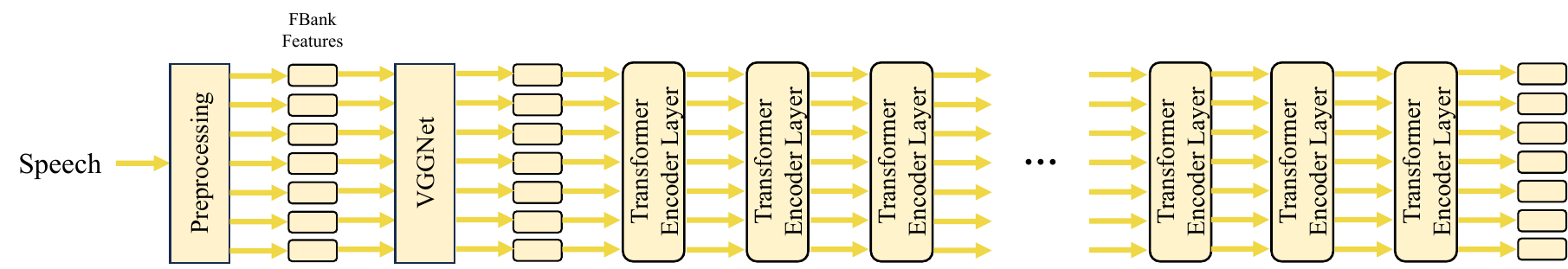}
  \caption{The structure of the speeh semantic encoder.}
  \label{fig:speech_encoder}
\end{figure}

Speech signals are inherently temporal sequences. It typically influenced by speaker-specific characteristics and environmental noise. To handle this complexity, we build speech semantic encoder by combining Convolutional Neural Network (CNN) and Transformer\cite{vaswani2017attention}, which is depicted in \fref{fig:speech_encoder}. 
 
In the preprocessing stage, we first extract the FBank (Filter Bank) features of the speech data, which can retain the key spectrum information of the speech signal. In order to enhance the ability of the encoder to extract speech information, many studies introduce CNN for further feature extraction. This is because one of the keys to improving the performance of speech tasks is to overcome the diversity of speech signals. The FBank features still remain this diversity. CNN has spatio and temporal translation invariance. Applying CNN to acoustic modeling can overcome the diversity. Therefore, we design the speech semantic encoder based on VGGNet\cite{simonyan2014very} for further feature extraction, which is widely used in speech domain. Moreover, we replace the original convolution with causal convolution\cite{van2016wavenet} for better processing temporal characteristics. Different from origin convolution, the core idea of causal convolution is that the output at the current time point can only depend on the current and past inputs. This ensures the causality of the signal processing. At the same time, this can also better capture the time-varying information in speech data. The entire process can transform raw speech data $D^\mathcal{S}$ into feature map $F^\mathcal{S}_{vgg} \in \mathbb{R}^{L_{\mathcal{S}} \times P}$. $L_\mathcal{S}$ and $P$ denote the length and width of the feature map, respectively.

The application of CNN advances feature extraction from the spectrum of speech signals. However, for speech data with temporal properties, features extracted by CNN are shallow features. To bridge this gap, we add another Transformer-based part to the speech semantic encoder. The architecture of Transformer is suitable for processing problems with temporal characteristics and can extract higher-level semantic features. We treat the feature map as sequence embeddings. $L_\mathcal{S}$ and $P$ are seen as the length of the sequence and the dimension of the embedding vector, respectively. The feature map is sent to multiple transformer encoder layers for further feature extraction, and the semantic features $F^{\mathcal{S}} \in \mathbb{R}^{L_{\mathcal{S}} \times P}$ can be obtained.

\subsubsection{Video Semantic Encoder}
Video modal data can be viewed as a collection of images, which are stored according to time. It means that video modal data has temporal characteristics. The difficulty in video modal data also lies in processing the temporal characteristics.

\begin{figure}[htbp]  
  \centering
  \includegraphics[scale=0.22]{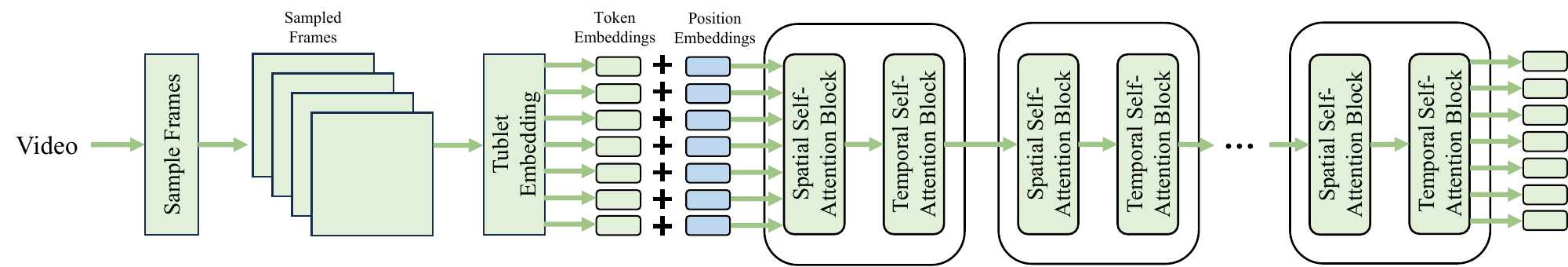}
  \caption{The structure of the video semantic encoder.}
  \label{fig:video_encoder}
\end{figure}

To solve this problem, we build our video semantic encoder based on Video Vision Transformer (ViViT)\cite{arnab2021vivit}, which is shown in \fref{fig:video_encoder}. ViViT is a model based on Transformer architecture, which is good at capturing rich semantic information from videos. During preprocessing, we sample frames from video data and apply data augmentation techniques to each frame. The input data can be denoted as $D^\mathcal{V} \in \mathbb{R}^{N_F \times C \times H \times W}$, where $N_F$, $C$, $H$ and $W$ are the number of sampled frames, the number of channels, width and height, respectively. Our video semantic encoder first needs to embed $D^\mathcal{V}$ into token embeddings and we adopt tubelet embedding\cite{arnab2021vivit} here. Each time we take a small patch along the length and width of a frame, and take $N_f-1$ patches located in the same region of subsequent $N_f-1$ frames along the time dimension. All patches are merged together to form a tube. All tubes have no overlapping parts. For the tube of dimension $N_f \times h \times w$, the total count $L_\mathcal{V}$ is given by $L_\mathcal{V} = n_f \times n_h \times n_w$, 
where $n_f={N_F}/{N_f}$, $n_h={H}/{h}$, and $n_w={W}/{w}$. Then all tubes are projected into $P$-dimensional token embeddings 
$E^\mathcal{V}_{tok} \in \mathbb{R}^{L_{\mathcal{V}} \times P}$. For token embeddings, we incorporate position embeddings $E^\mathcal{V}_p \in \mathbb{R}^{L_{\mathcal{V}} \times P}$ to introduce sequence position information, which is given by
\begin{equation}
  E^\mathcal{V}_{in}  = E^\mathcal{V}_{tok} + E^\mathcal{V}_p
\end{equation}
where $E^\mathcal{V}_{in} \in \mathbb{R}^{L_{\mathcal{V}} \times P}$ is the input embeddings. These embeddings are all learnable parameters.

The input embeddings $E^\mathcal{V}_{in}$ combine both spatial and temporal information. To better extract semantic features from them, our video semantic encoder decomposes embeddings into two dimensions for processing: space and time. In the video semantic encoder, each layer is composed of two self-attention blocks, namely a spatial self-attention block and a temporal self-attention block. 

Note that the input embeddings are from token embeddings and token embeddings are linearly mapped from tubes. It means that some input embeddings are from the same time series while some are from the same spatial region. In each layer, input embeddings are first fed into a spatial self-attention block. Only input embeddings from the same spatial region will participate in the operations in the spatial self-attention block, and those from different spatial regions will not compute attention for each other. Then we can derive the spatial features $F^{\mathcal{V}}_s \in \mathbb{R}^{L_{\mathcal{V}} \times D}$. $F^{\mathcal{V}}_s$ is sent to the temporal self-attention block, and the temporal attention is calculated for all input embeddings from the same time series. Then the temporal features $F^{\mathcal{V}}_t \in \mathbb{R}^{L_{\mathcal{V}} \times D}$ can be acquired. Through such alternating extraction of spatial and temporal features, we are ultimately able to obtain the semantic features $F^{\mathcal{V}} \in \mathbb{R}^{L_{\mathcal{V} \times D}}$ of video modal data. This method fuses spatial and temporal features, which is more beneficial for tasks in the video modality.

To prevent negative transfer among unrelated or even conflicting tasks, we introduce task embeddings\cite{achille2019task2vec} for each task to enhance the differentiation among tasks. We number the tasks sequentially and embed serial numbers into $P$-dimensional vectors, which are denoted as $E^{T}_i \in \mathbb{R}^P$. $i$ represents the task serial number and $i \in \mathcal{K}$. For semantic features of task $T_i$, we concat $E^{T}_i$ and them together. This design can reduce negative interference among tasks and improve learning outcomes.

\subsection{Fusion Module Design}
\begin{figure}[htbp]
	\centering
	\begin{minipage}{\linewidth}
    \begin{subfigure}[b]{1\textwidth}
		\centering
		\includegraphics[width=0.85\linewidth]{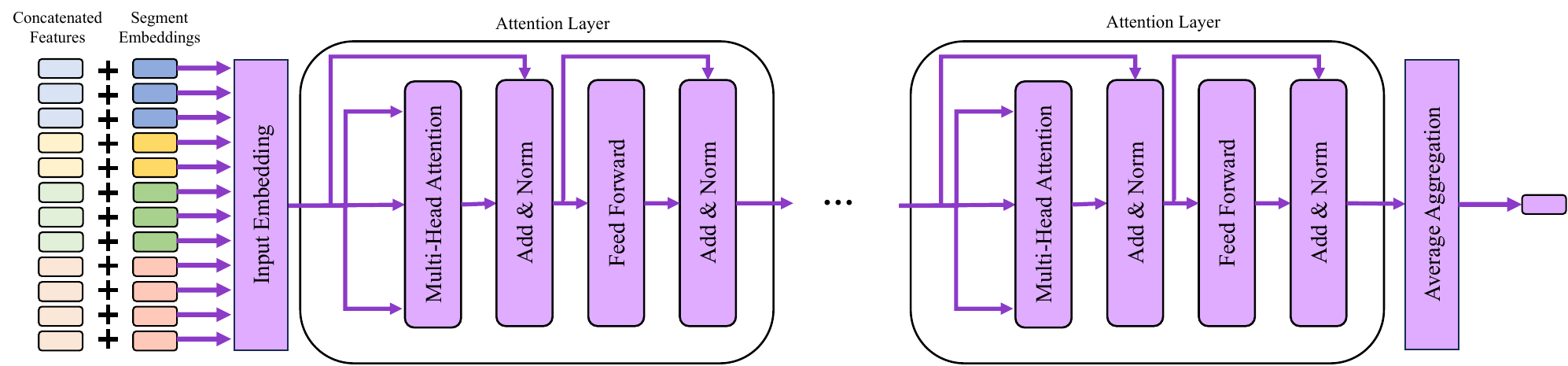}
    \caption{{\fontsize{8pt}{10pt}\selectfont The structure of the fusion module.}}
  \end{subfigure}
	\end{minipage}
	\qquad

	\begin{minipage}{\linewidth}
    \begin{subfigure}[b]{0.45\textwidth}
      \centering
      \includegraphics[width=\textwidth]{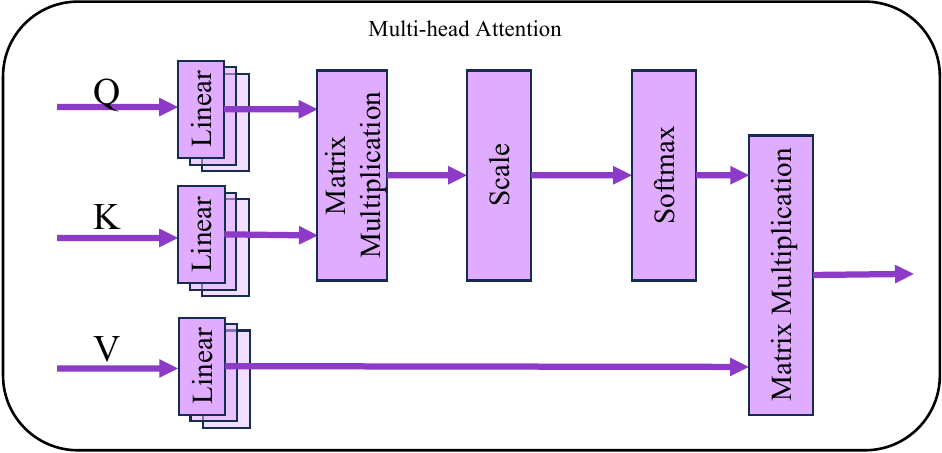}
      \caption{{\fontsize{8pt}{10pt}\selectfont The structure of the multi-head self-attention.}}
  \end{subfigure}
  \begin{subfigure}[b]{0.45\textwidth}
      \centering
      \includegraphics[width=\textwidth]{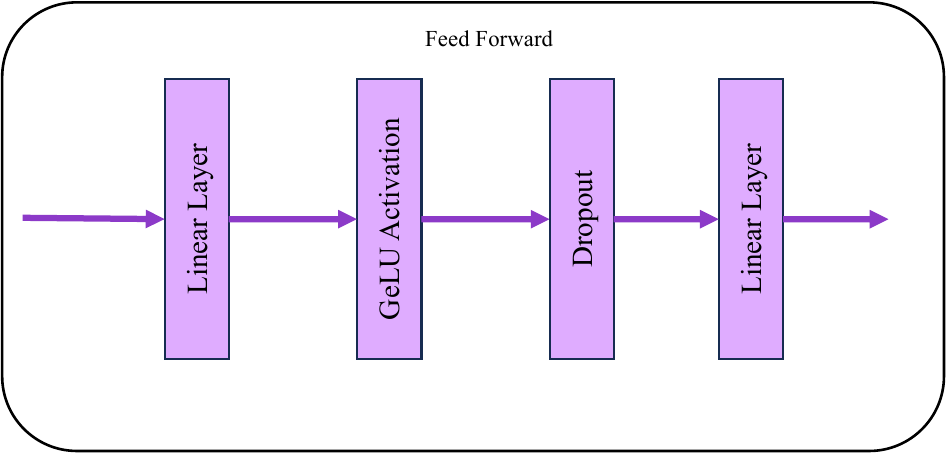}
      \caption{{\fontsize{8pt}{10pt}\selectfont The structure of the feedforward neural network.}}
  \end{subfigure}
	\end{minipage}
  \caption{This module can capture the complementary information and reduce redundancy.}
  \label{fig:fusion_module}
\end{figure}

The semantic encoder we design for each modality takes into account the characteristics of each modal data. It aims at extracting distinct semantic features. For multi-modal tasks, semantic features of different modalities provide different details, and fusing them can improve the performance of the model while reducing redundancy. However, due to the heterogeneity of different modal data, these semantic features are not in the same semantic space. This means that neglecting multi-modal fusion and simply concatting them together will cause misalignment of semantic information. To solve this problem, we design a fusion module based on BERT\cite{devlin2018bert}, which is
depicted in \fref{fig:fusion_module}. The key of this method is to use the self-attention mechanism to achieve fusion of multi-modal semantic features. The self-attention mechanism is adept at capturing the interdependencies among different modalities. It dynamically weights the importance of each modal semantic features, effectively realigning them into a unified semantic space. This enables the model to leverage complementary information.

Assume that $\mathcal{D}=\{(D^{M_1}, D^{M_2}, ..., D^{M_m}, Y)\}$ is a multi-modal dataset, where $D^{M_m}$ and $Y$ are the source data with $M_m$ modality and its corresponding labels, respectively. Input the data into the corresponding semantic encoder, we can obtain the corresponding semantic features. These semantic features are represented as $F^{M_1} \in \mathbb{R}^{L_{M_1} \times P}, F^{M_2} \in \mathbb{R}^{L_{M_2} \times P}, ..., F^{M_m} \in \mathbb{R}^{L_{M_m} \times P}$. In this case, we regard the semantic features extracted from each modality as sequence features. $L_{M_j}$ is the length of the sequence and $P$ is the dimension of feature vecotors, where $j$ is between $1$ and $m$. Subsequently, these features are sent to the fusion module for fusion. The steps for our fusion module to fuse different modal semantic features are as follows.

\begin{itemize}
  \item \textit{Step 1 (Concatenation):} First, we need to concat semantic features of different modalities. We can get the concatenation of semantic features given by
  \begin{equation}
    F^C = \mbox{concat}(F^{M_1}, F^{M_2}, ..., F^{M_m}) 
  \end{equation}
  where $F^C \in \mathbb{R}^{L' \times P}$ and $L' = \sum_{j=1}^{m}L_{M_j}$.
  \item \textit{Step 2 (Segment Embeddings):} Unlike the traditional Transformer models, our fusion module does not incorporate position embeddings. This is because position embeddings have been added in the text, speech, and video semantic encoder. And image semantic features are essentially the feature map, so there is no need for position embeddings. Moreover, for the concatenated features, we think they are all equally important and the sequence order should have no effect on the task performance. Therefore, we only consider segment embeddings. In this context, we treat different modalities of semantic features as distinct segments. Each modal $M_j$ has a unique segment embedding vector $E_s^{M_j} \in \mathbb{R}^{1 \times P}$. For each sequence vector in $F^C$, we select the corresponding segment embedding vector. All segment embedding vectors are concatenated together to get the segment embeddings $E_s \in \mathbb{R}^{L' \times P}$. Then, the input embeddings are denoted as
  \begin{equation}
    F^{in} = F^C + E_s, \quad F^{in}  \in \mathbb{R}^{L' \times P}.
  \end{equation}
  Through segment embeddings, the distinction of semantic features of different modalities can be enhanced, allowing our fusion module to better capture the distinct semantic features of each modality and cross-modal interdependencies. In addition to segment embeddings, we concat the task embedding vector and $F^{in}$ to prevent negative transfer of effects. Therefore, the dimension of $F^{in}$ are $\mathbb{R}^{L \times P}$, where $L=L'+1$.

  \item \textit{Step 3 (Attention Layer):} $F^{in}$ is then input into multiple attention layers. We design 6 attention layers in our fusion module. Each layer consists of two parts: multi-head self-attention (MSA) and feedforward neural network (FFN). Suppose that $H^{(l)} \in \mathbb{R}^{L \times P}$ is the output by the $l$-th attention layer, and $H^{(0)}$ is $F^{in}$. Then $H^{(l)}$ can be denoted as
  \begin{equation}
    \begin{aligned}
      H^{(l)'} & = \mbox{LN}(\text{MSA}(H^{(l-1)}) + H^{(l-1)}) \\
      H^{(l)} & = \mbox{LN}(\text{FFN}(H^{(l)'}) + H^{(l)'}) \\
    \end{aligned}
  \end{equation}
  where LN is the layer normalization and $H^{(l)'}$ is the intermediate features of $l$-th layer that passes through MSA. MSA mechanism allows the fusion module to focus on different parts of input sequences, enabling it to capture a wide range of dependencies. This is given by
  \begin{equation}
    \text{MSA}(H^{(l-1)}) = \text{concat}(\text{head}_1, ..., \text{head}_h)W^O.
  \end{equation}
  $W^O \in \mathbb{R}^{P \times P}$ is a learnable parameter matrix and $\text{head}_i \in \mathbb{R}^{L \times P_h}$ is calculated by self-attention (SA). $P_h$ is the dimension of each head. In our work, we consider 12 heads, and $P_h = P / 12$. For each head output by SA, it can be represented as
  \begin{equation}
    \begin{aligned}
      \text{head}_i & =  \text{SA}(Q, K, V) \\
                    & = \text{softmax}(\frac{(QW_i^Q)(KW_i^K)^T}{\sqrt{P}}) VW_i^V \\
    \end{aligned}
  \end{equation} 
where $W_i^Q \in \mathbb{R}^{P \times P_h}$, $W_i^K \in \mathbb{R}^{P \times P_h}$, $W_i^V \in \mathbb{R}^{P \times P_h}$ are weight matrices for the query $Q \in \mathbb{R}^{L \times P}$, key $K \in \mathbb{R}^{L \times P}$ and value $V \in \mathbb{R}^{L \times P}$, respectively. Since we use self-attention mechanism, $Q$, $K$, $V$ are actual is $H^{(l-1)}$. In softmax, we added a scaling factor $\sqrt{P}$, which is to prevent the gradient vanishing problem caused by excessive attention value. The output of MSA is then processed by the FFN, which applies two linear layers with the GeLU activation function:
\begin{equation}
  \mbox{FFN}(H^{(l)'}) = \mbox{GeLU}(H^{(l)'}W_1 + b_1)W_2 + b_2
\end{equation}
where $W_1, b_1, W_2, b_2$ are parameters of the two linear layers.
\item \textit{Step 4 (Average Aggregation):} 
After six attention layers, we derive the features $H^{(6)} \in \mathbb{R}^{L \times P}$. Then, we perform an average aggregation on the first dimension to obtain a fused feature vector $F^f \in \mathbb{R}^{1 \times P}$. The fused semantic features then are sent to channel encoder. For single-modal tasks, the semantic features do not go through the fusion module, but are directly dispatched to the channel encoder.
\end{itemize}

Semantic features from differenct modalities often face misalignment issues in multi-modal fusion due to discrepancies in semantic spaces. Employing the attention mechanism enables the fusion module to align these varied semantic features, enhancing multi-modal task performance while reducing transmission redundancy.

\subsection{Channel Encoder}
\begin{figure}[htbp]  
  \centering
  \includegraphics[width=0.8\linewidth,scale=1.5]{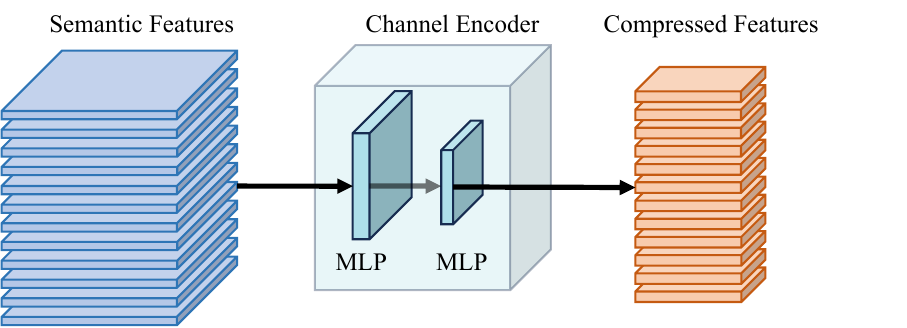}
  \caption{The structure of the channel encoder.}
  \label{channel_encoder}
\end{figure}

In the channel encoder, we use two layers of Multilayer Perceptron (MLP), which is shown in \fref{channel_encoder}. The first MLP layer is used to receive the original semantic features and compress them. The second MLP layer is used to further compress the output of the first layer. This layer has fewer neurons, resulting in a lower output dimension compared to the first layer. In this way, the network can learn more abstract feature representations. The output of the second MLP layer can be considered a compact representation of the original semantic features, with higher information density, while further reducing communication overhead of semantic communication.

\subsection{Channel Decoder}
\begin{figure}[htbp]  
  \centering
  \includegraphics[width=0.8\linewidth,scale=1.5]{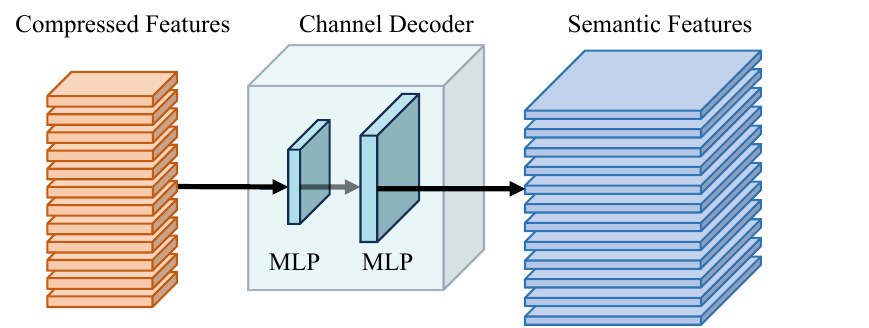}
  \caption{The structure of the channel decoder.}
  \label{channel_decoder}
\end{figure}

The compressed semantic features contain the core information of the original features but have lower dimensions. To restore the information, we also use two layers of MLP for channel decoding, which is depicted in \fref{channel_decoder}. The first MLP layer restores the information of the original features. The second MLP layer is used to further improve the decoding process. This layer has more neurons, enabling it to learn more complex feature reconstruction patterns.

\subsection{Semantic Decoder}
\begin{figure}[htbp]  
  \centering
  \includegraphics[width=0.8\linewidth,scale=1]{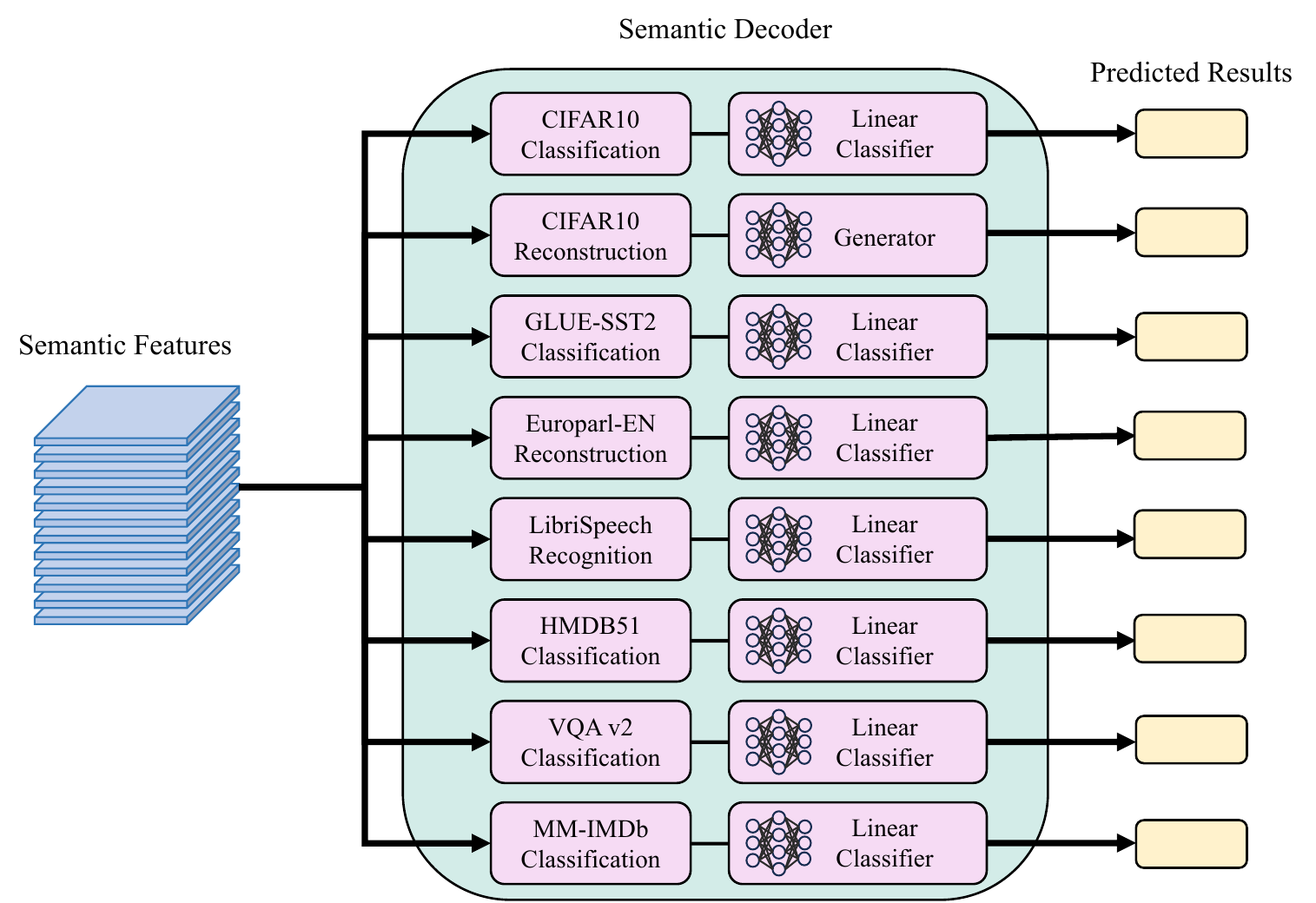}
  \caption{The structure of the semantic decoder.}
  \label{semantic_decoder}
\end{figure}

We choose a simplified and efficient approach, avoiding complex neural networks as  the semantic decoder. It is demonstrated in \fref{semantic_decoder}. We design specific lite neural networks for each task head which only contains one or two linear layers. This simplified design is based on the following considerations: First, we find that a complex architecture does not necessarily bring significant improvements in performance. Instead, using simple linear layers, sufficient accuracy requirements can be achieved. Furthermore, the simple linear layer architecture allows us to significantly reduce the consumption of computing resources without sacrificing too much accuracy. This design makes our framework excel in flexibility.

\subsection{Training and Testing}
We conduct joint training for all tasks. Each time, we randomly select one task for training. During the training process, we employ the Adam\cite{kingma2014adam} optimizer for parameter optimization. Specifically, we create a separate Adam optimizer for each task. This allows the model to adjust the learning rate and parameter updates for each task more effectively. The whole process is shown in \aref{training}. During this process, the semantic encoder $S(\cdot; \alpha^{M_j}_{T})$ of each modality $M_j$, fusion module $F(\cdot; \gamma)$, channel encoder $C(\cdot; \beta_{T})$, channel decoder $C^{-1}(\cdot; \beta_{R})$, and semantic decoder $S^{-1}(\cdot; \beta_{R})$ will be optimized. We evaluate the performance of the previously trained model in AWGN channel and Rayleigh fading channel under various SNR (signal-to-noise) conditions. The whole process is shown in \aref{testing}.

\begin{algorithm}
	\renewcommand{\algorithmicrequire}{\textbf{Input:}}
	\renewcommand{\algorithmicensure}{\textbf{Output:}}
	\caption{Training Stage}
	\label{training}
	\begin{algorithmic}[1]
    \REQUIRE Task set $T$
		\STATE Initialization: Initialize parameters $\alpha_\mathcal{T}$, $\beta_\mathcal{T}$, $\alpha_\mathcal{R}$ and $\beta_\mathcal{R}$.
    \STATE Initialization: Initialize Adam optimizers $optim_i$ for each task.
    \FOR {i $\leftarrow$ 1 to $N$}
    \STATE Randomly choose one task $T_i$ from $T$. \\
    \STATE Generate a batch of samples $\mathcal{D}_i$ and corresponding labels $Y_i$ from the remaining unselected data of the selected task. \\ 
    \IF {task $T_i$ has no remaining data} 
    \STATE Remove $T_i$ from task sets $T$.
    \ENDIF
    \IF {$T$ is a single-modal task and the modality is $M$} 
      \STATE $ X_i \leftarrow C(S(D_i; \alpha^M_{T}); \beta^M_{T})$ \\
    \ELSE
      \FOR {each modality $M_j$}
        \STATE $F^{M_j} \leftarrow S(D^{M_j}_i; \alpha^{M_j}_{T})$
      \ENDFOR
      \STATE $X_i \leftarrow C(F((F^{M_1}, F^{M_2}, ..., F^{M_m}); \gamma); \beta_{T})$
    \ENDIF
    \STATE $\hat{X_i} \leftarrow HX_i + N$ 
    \STATE $\hat{Y_i} \leftarrow S^{-1}(C^{-1}(X_i; \beta_{\mathcal{R}}); \alpha_{\mathcal{R}})$ \\ 
    \STATE $\mbox{loss} \leftarrow L_i(\hat{Y_i}, Y_i)$
    \STATE Update parameters by using $optim_i$.
    \ENDFOR
    \ENSURE Trained networks, $S(\cdot; \alpha_{T})$, $S^{-1}(\cdot; \alpha_{R})$, $C(\cdot; \beta_{T})$, $C^{-1}(\cdot; \beta_{R})$ and $F(\cdot; \gamma)$.
	\end{algorithmic}  
\end{algorithm}

\begin{algorithm}
	\renewcommand{\algorithmicrequire}{\textbf{Input:}}
	\renewcommand{\algorithmicensure}{\textbf{Output:}}
	\caption{Testing Stage}
	\label{testing}
	\begin{algorithmic}[1]
    \REQUIRE Task set $T$, trained networks, and a wide range of SNR values 
    \FOR {each task $T_i$}
      \FOR {each SNR value $snr$}
      \STATE Generate Gaussian noise $N$ under the $snr$.
      \STATE $S \leftarrow$ All data batches of $T_i$.
      \IF {$T$ is a single-modal task and the modality is $M$} 
      \STATE $ X_i \leftarrow C(S(D_i; \alpha^M_{T}); \beta^M_{T})$ \\
    \ELSE
      \FOR {each modality $M_j$}
        \STATE $F^{M_j} \leftarrow S(D^{M_j}_i; \alpha^{M_j}_{T})$
      \ENDFOR
      \STATE $X_i \leftarrow C(F((F^{M_1}, F^{M_2}, ..., F^{M_m}); \gamma); \beta_{T})$
    \ENDIF
    \STATE $\hat{X_i} \leftarrow HX_i + N$ 
    \STATE $\hat{Y_i} \leftarrow S^{-1}(C^{-1}(X_i; \beta_{\mathcal{R}}); \alpha_{\mathcal{R}})$ \\ 

      \STATE $p_i^{snr} \leftarrow P_i(\hat{Y_i}, Y_i)$

      \ENDFOR
    \ENDFOR
    \ENSURE The performance of different tasks.
	\end{algorithmic}  
\end{algorithm}

\section{Simulation Results}
\label{results}

In this section, we compare the performance of the proposed MFMSC with several benchmarks, MMSC, T-DeepSC and U-DeepSC, traditional methods under the AWGN and Rayleigh fading channels.

\subsection{Simulation Settings}
To avoid the conflits of different tasks, we set an Adam optimizer for each task. Except for the speech recognition task, where the learning rate is 2e-4, the learning rates of other tasks are set to 1e-4. To better verify the effectiveness of MFMSC, we include multiple baselines and existing models for comparison. 

\subsubsection{MFSC} 
We construct an MFSC (multi-modal fusion-based semantic communication) framework for comprision of multi-task capabilities. The architecture of this framework is exactly the same as MFMSC, whereas the difference lies in the training and testing process. MFMSC employs a joint training, and the performance is tested after training is completed. Whereas MFSC uses independent training and testing for each task.

\subsubsection{MMSC} 
To evaluate the effectiveness of the fusion module in MFMSC, we introduce the MMSC (multi-modal multi-task semantic communication) framework. MMSC differs from our proposed MFMSC in its semantic encoder design. MMSC retains the semantic encoders of each modality in MFMSC while removing the fusion module. For single-modal tasks, the raw data is still extracted through the corresponding semantic encoder to extract semantic features in MMSC. This process is consistent with MFMSC. For multi-modal tasks, different modal data is sent to the corresponding semantic encoder to extract semantic features. Then, MMSC concats the semantic features together and there is no fusion module. The concatenation of semantic features is input into the channel encoder. The processing on the receiver side of the MMSC model is the same as MFMSC. In addition, the training strategy of MMSC is still joint training.
\subsubsection{U-DeepSC} 
U-DeepSC\cite{zhang2022unified} is a unified multi-task multi-modal semantic communication framework proposed by Zhang \emph{et al}. It is based on Transformer and supports three modal data of image, text and speech. Video modal data is treated as multi-modal data of these three modalities. It should be noted that the multi-modal processing in U-DeepSC is similar to MMSC, which is direct concatenation without a fusion module.
\begin{figure*}[htbp]
  \centering
  \begin{subfigure}[b]{0.21\textwidth}
      \centering
      \includegraphics[width=\textwidth]{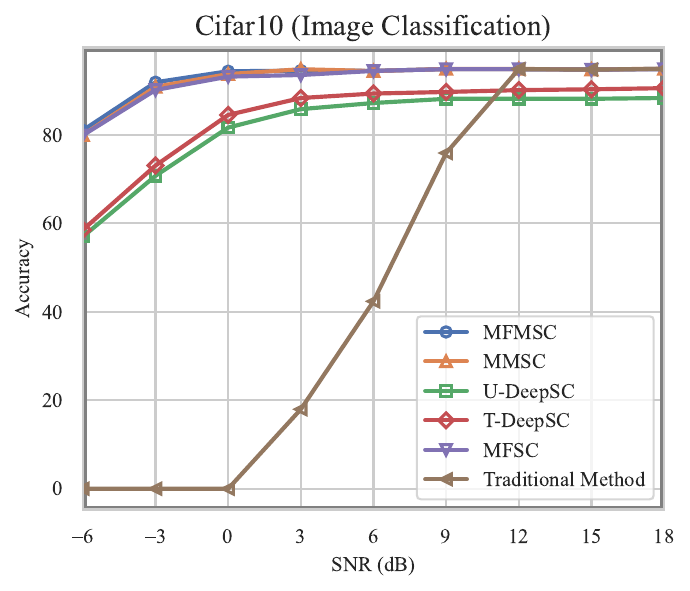}
      \caption{{\fontsize{6pt}{10pt}\selectfont Image Classification Task}}
  \end{subfigure}
  \begin{subfigure}[b]{0.21\textwidth}
      \centering
      \includegraphics[width=\textwidth]{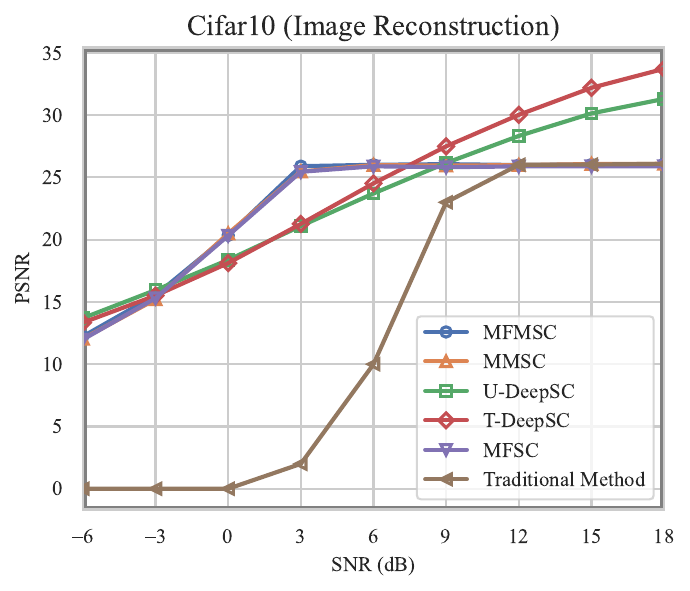}
      \caption{{\fontsize{6pt}{10pt}\selectfont Image Reconstruction Task}}
  \end{subfigure}
  \begin{subfigure}[b]{0.21\textwidth}
      \centering
      \includegraphics[width=\textwidth]{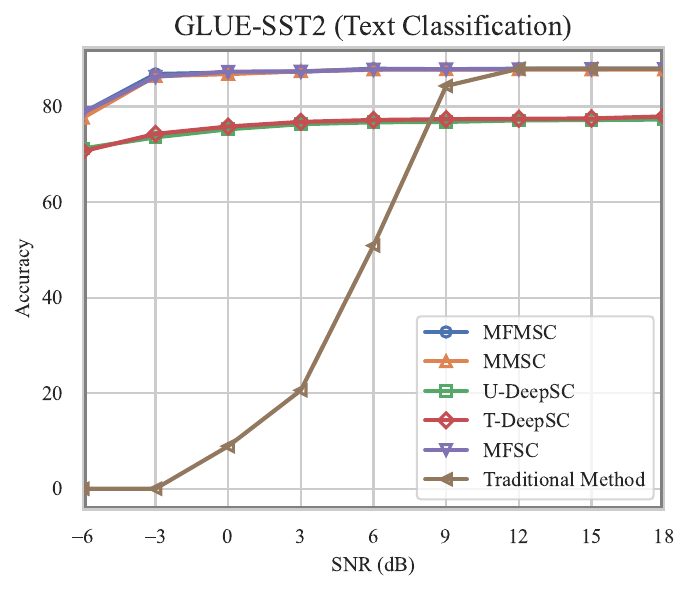}
      \caption{{\fontsize{6pt}{10pt}\selectfont Sentiment Analysis Task}}
  \end{subfigure}
  \begin{subfigure}[b]{0.21\textwidth}
      \centering
      \includegraphics[width=\textwidth]{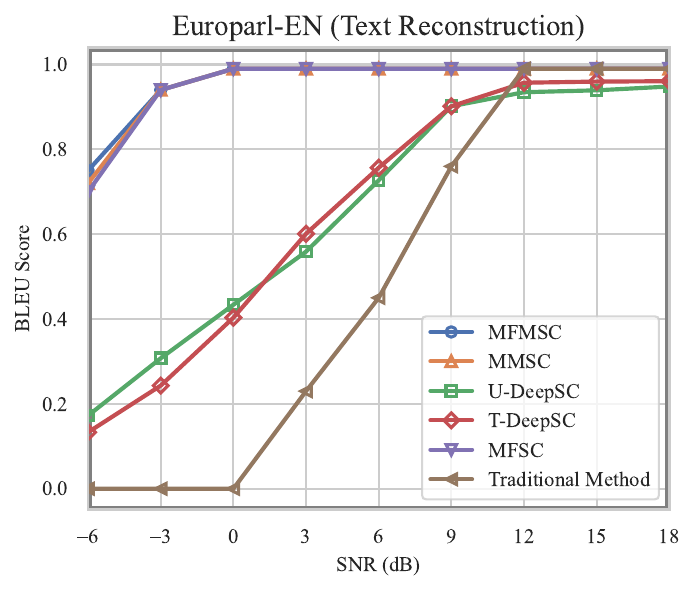}
      \caption{{\fontsize{6pt}{10pt}\selectfont Text Reconstruction Task}}
  \end{subfigure}

\begin{subfigure}[b]{0.21\textwidth}
    \centering
    \includegraphics[width=\textwidth]{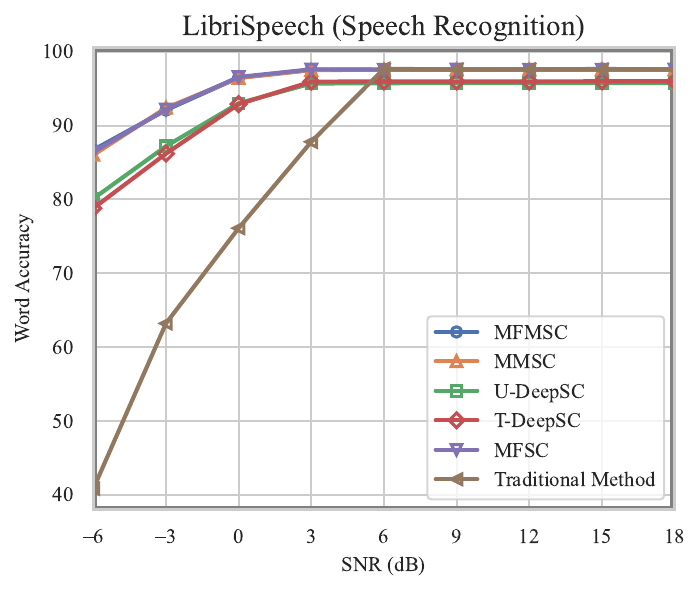}
    \caption{{\fontsize{6pt}{10pt}\selectfont Speech Recognition Task}}
\end{subfigure}
\begin{subfigure}[b]{0.21\textwidth}
    \centering
    \includegraphics[width=\textwidth]{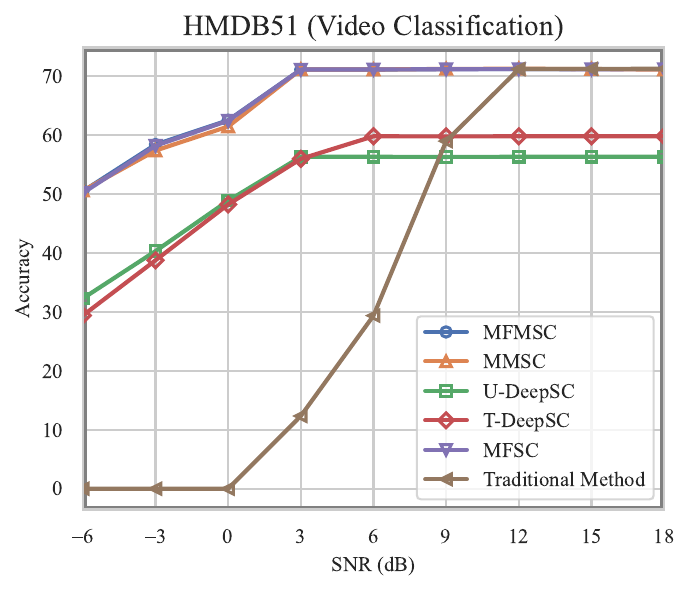}
    \caption{{\fontsize{6pt}{10pt}\selectfont Video Classification Task}}
\end{subfigure}
\begin{subfigure}[b]{0.21\textwidth}
    \centering
    \includegraphics[width=\textwidth]{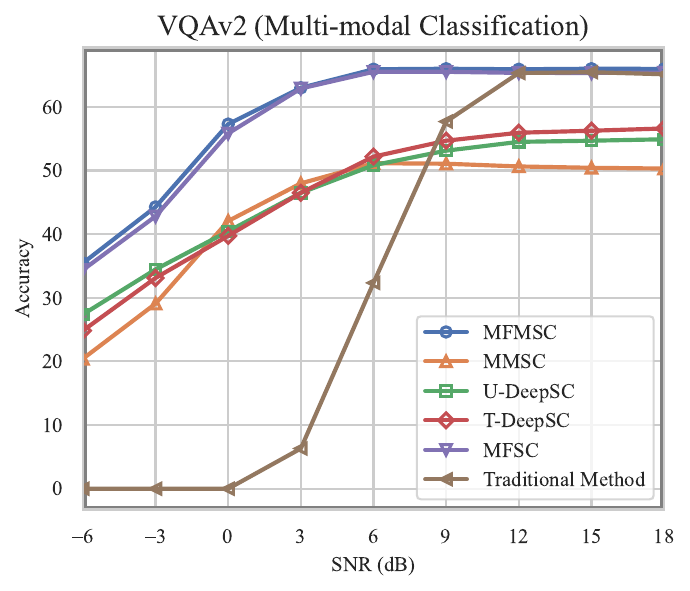}
    \caption{{\fontsize{6pt}{10pt}\selectfont Visual Question Answering Task}}
    \label{fig:awgn_vqa}
\end{subfigure}
\begin{subfigure}[b]{0.21\textwidth}
    \centering
    \includegraphics[width=\textwidth]{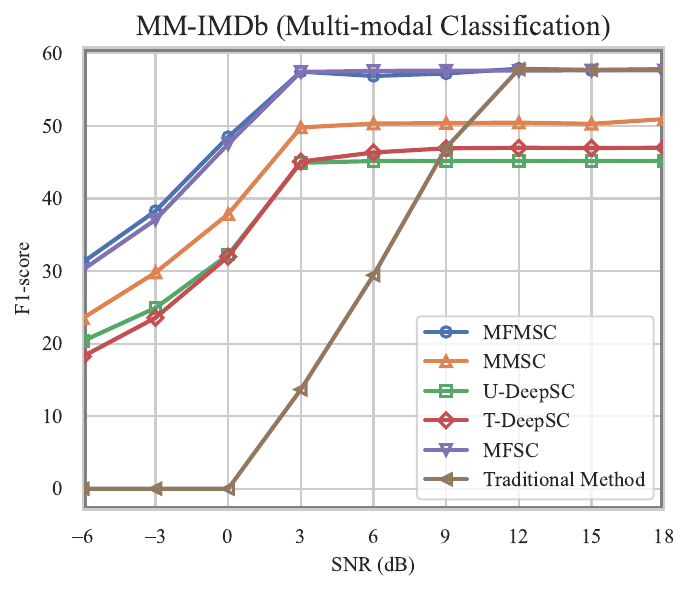}
    \caption{{\fontsize{6pt}{10pt}\selectfont Genre Classification Task}}
    \label{fig:awgn_imdb}
\end{subfigure}
\caption{Performance comparison in AWGN channel under different SNR environments.}
\label{fig:performance_awgn}
\end{figure*}
\subsubsection{T-DeepSC} 
The architecture of T-DeepSC\cite{zhang2022unified} and U-DeepSC is the same. The difference between them is the training process. T-DeepSC is trained independently for each task and finally multiple models are saved, while U-DeepSC is trained jointly.

\subsubsection{Traditional Methods} 
This is the traditional separate source-channel coding. For different modal data, we adopt different communication encoding methods. For image data, we use Joint Photographic Experts Group (JPEG) and Low Density Parity Check Code (LDPC) as image source encoding and image channel encoding, respectively. For video data, we adopt the H.264 video compression codec for source encoding. For text data, we use 8-bit Unicode Transform Format (UTF-8) encoding and Turbo encoding as text source encoding and text channel encoding, respectively. For speech signals, 16-bit pulse code modulation (PCM) and LDPC are used as source coding and channel coding, respectively.

\subsection{Comparison of Task Performance}
\begin{figure}[htbp]
  \centering
  \begin{subfigure}[b]{0.21\textwidth}
      \centering
      \includegraphics[width=\textwidth]{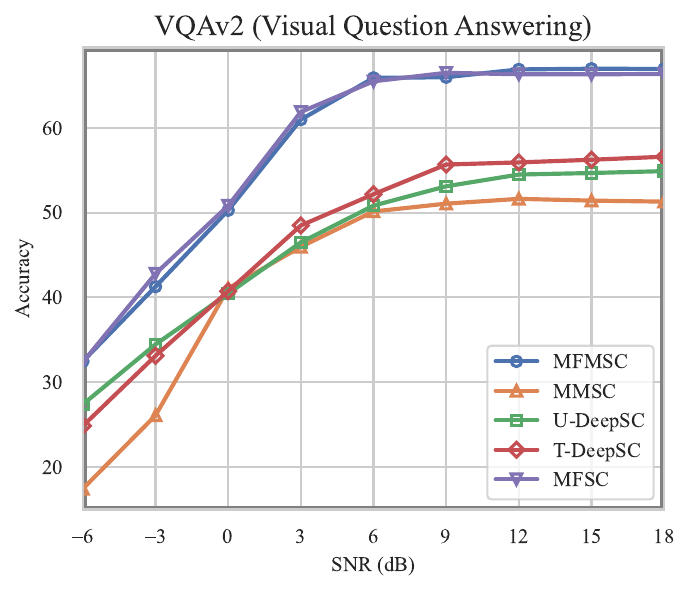}
      \caption{{\fontsize{6pt}{10pt}\selectfont Visual Question Answering Task}}
      \label{fig:rayleigh_vqa}
  \end{subfigure}
  \begin{subfigure}[b]{0.21\textwidth}
      \centering
      \includegraphics[width=\textwidth]{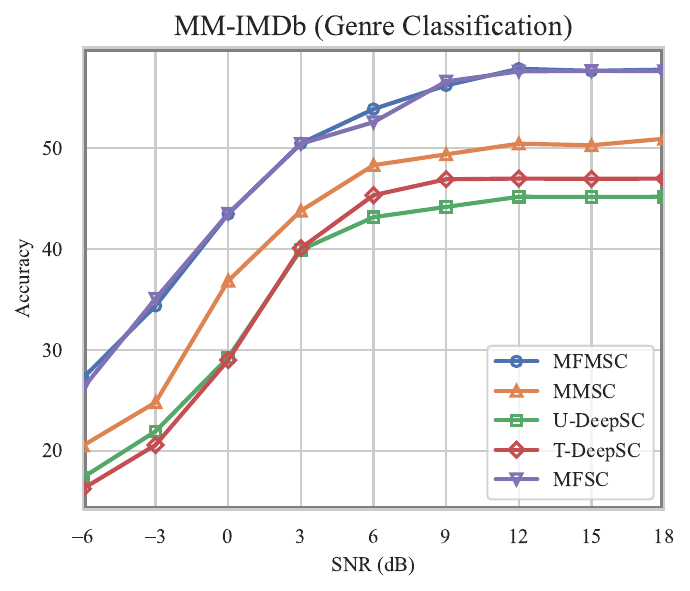}
      \caption{{\fontsize{6pt}{10pt}\selectfont Genre Classification Task}}
      \label{fig:rayleigh_imdb}
  \end{subfigure}
  \caption{Performance comparison of in Rayleigh fading channel under different SNR environments.}
  \label{fig:performance_rayleigh}
\end{figure}

We conduct extensive evaluation on multiple tasks. \fref{fig:performance_awgn} illustrates the performance of our proposed framework under AWGN channel. Traditional communication methods perform poorly under low SNR conditions, while those semantic communication models show good performance. Each of them demonstrates the ability of maintaining the semantic integrity against the backdrop of noise. As the SNR increases to higher levels, the MFMSC stands out, often achieving the highest performance. Furthermore, the performance of our proposed MFMSC is close to MFSC. This shows that MFMSC has good support for multi-task semantic communication systems. Moreover, we extend the test to Rayleigh fading channel conditions, which is depicted in \fref{fig:performance_rayleigh}. Since we pay more attention to multi-modal tasks, only the performance of VQA v2 and MM-IMDb is shown in this figure. MFMSC once again demonstrates strong performance, showing no significant performance degradation. The findings indicate that our framework effectively extracts the semantic information, showcasing its application potential.

\subsection{Effectiveness of Fusion Module}
In order to better compare the effect of our multi-modal fusion, we set up a baseline model MMSC. Our experimental results are demonstrated in the \fref{fig:awgn_vqa}, \fref{fig:awgn_imdb}, \fref{fig:rayleigh_vqa} and \fref{fig:rayleigh_imdb}. Whether it is the AWGN channel or the Rayleigh fading channel, the performance of MFMSC is significantly stronger than that of MMSC. This is beacuse that MMSC lacks the fusion module and often fail to capture the complementary features of different modalities. The drawback results in reduced task performance and higher communication overhead. Compared with the MMSC, MFMSC successfully achieves data fusion among different modalities through the fusion module. In addition, for single-modal tasks, we can find that the fusion module has almost no impact on the performance of them. From \fref{fig:performance_awgn}, it is apparent that the performance of MFMSC and MMSC is essentially on par in single-modal tasks. We extend our study analysis to T-DeepSC and U-DeepSC. These two models that similarly do not take advantage of modal fusion in their architectures. Like MMSC, U-DeepSC and T-DeepSC concat multi-modal semantic features together and then transmit them into the physical channel. We find that their performance is also significantly inferior to our MFMSC model. Clearly, the lack of fusion mechanism limits their ability to exploit complementary features. Our experiments demonstrate that MFMSC greatly enhances multi-modal task outcomes. And this is due to our BERT-based fusion module. This module uses the self-attention mechanism to align and fuse the semantic features in different semantic spaces of each modality, thereby realizing the interaction of modal information and making full use of the complementary information among modalities.

\subsection{Comparison of Communication Overhead}
For semantic communication, the evaluation of task performance is important, but equally important is the communication overhead. In practical applications, especially when resources are limited, communication overhead often determine whether a system has practical value. Considering that multi-modal tasks will bring additional communication overhead, in this work, we compare the communication overhead between the MFMSC and other models in multi-modal tasks.

In MFMSC, the fusion module plays a great role in eliminating redundancy. For the multi-modal data, each modality $M_j$ is extracted semantic features through a semantic encoder and the dimension is represented as $\mathbb{R}^{{L_{M_j}} \times P}$. ${L_{M_j}}$ represents the sequence length of modal $M_j$, and $P$ represents the feature dimension. Then we concat them and task embedding vecotor together. The concatenation is sent to the fusion module. Assuming that there is $m$ modalities in total, then the dimension of concatenated semantic features is $\mathbb{R}^{L \times P}$, where $L = \sum_{j=1}^{m}L_{M_j} + 1$. We fuse these modalities into a more compact representation, and the dimension of fused data is $\mathbb{R}^{1 \times P}$. Therefore, compared with MMSC, we can reduce the amount of transmitted data to $1/L$ of the original amount through multi-modal fusion.

\begin{table}[!htbp]
  \caption{Comparison of communication overhead.} 
  \label{tab:costs}
  \centering
  \begin{tabular}{ccc} 
    \toprule 
      Model & VQA v2 & MM-IMDb   \\
      \midrule
      MMSC & 6.400 KB & 3.968 KB \\
      TDeepSC & 0.240 KB & 0.352 KB \\
      UDeepSC & 0.240 KB & 0.352 KB \\
      \textbf{MFMSC} & \textbf{0.128 KB} & \textbf{0.128 KB} \\
    \bottomrule 
  \end{tabular}
\end{table}

We calculate the amount of data transmitted per task instance in the VQA v2 and MM-IMDb datasets, which is shown in \tref{tab:costs}. Compared with MMSC, T-DeepSC and U-DeepSC, our MFMSC reduces the communication overhead by 98.0\%, 46.7\%, and 46.7\% on the VQA task, respectively. On the MM-IMDb task, the communication overhead are reduced by 96.8\%, 63.6\%, and 63.6\%, respectively. This shows that MFMSC can reduce the communication overhead while improving multi-modal task performance, giving it better application prospects.

\section{Conclusion}
\label{conclusion}
In this paper, we present a multi-modal fusion-based multi-task semantic communication framework. Unlike traditional methods, our model demonstrates excellent performance on multiple tasks and shows superiority on various evaluation metrics. Notably, when compared with other benchmarks, our method takes advantage of the complementarity among different modalities through multi-modal fusion. It significantly improves the performance of the model while effectively reducing redundant information. We successfully reduce the amount of data transmitted to $1/L$ of the original amount. Compared with U-DeepSC and T-DeepSC, our framework also provides better performance and lower communication overhead. This makes it more competitive in multi-task and multi-modal communication systems.


\bibliographystyle{IEEEtran}
\bibliography{ref}


\end{document}